# Estimation of the Base Pressure in Bluff Body Flows

Benjamin Bock [a]

[a]Tplus Engineering GmbH, Steinbeisstraße 25, 70771 Leinfelden-Echterdingen, Germany, Benjamin.Bock@tplus-engineering.de

**SUMMARY:**
A method for estimating the base pressure in bluff body flows is proposed. The estimation uses the velocity at the separation edge with the Bernoulli equation to compute the base pressure. The assumption is that this is the base pressure that will be approached when velocity fluctuations in the flow field are reduced. A constant velocity along the separating free streamline - as in two-dimensional bluff body wakes with a splitter plate - is a requirement to permit the application of the method. The suggested method is applied to a bluff body in a two-dimensional flow setup with and without a splitter plate. By reducing the velocity fluctuations, there are estimated increases in the base pressure of up to 58 % for the bluff body in two-dimensional flow without a splitter plate and up to 42 % with a splitter plate. The estimated base pressure of the two-dimensional flow without a splitter plate is compared to a case with active flow control. In this case, the velocity fluctuations caused by the Kármán vortex street are reduced. The comparison shows that the estimation of the base pressure can be achieved.

## 1. INTRODUCTION

Active flow control offers a promising approach for reducing the aerodynamic drag of bluff bodies. Due to the limit of energy sources, energy efficiency (and thus the constant reduction of aerodynamic drag) is an important topic. Active flow control uses actuators to change the flow, which can be used in a wide variety of ways and at different points. For example, the forces can be optimized to a target value via the external energy input of the actuator with an advantageous cost-benefit ratio. Particularly in aeronautics, progress has been made in recent years regarding the active flow control (Amitay et al., 1999; Clemons and Wlezien, 2016; Giorgi et al., 2015; Hosseini et al., 2018; Kral et al., 1997; Kral, 2000; Yang and Zha, 2017). Furthermore, there are effective approaches for bluff bodies, e.g., for two-dimensional flow (Bock et al., 2016; Chun and Sung, 1996; Dahan et al., 2012; Fujisawa and Takeda, 2003; Gong, 2015; Pastoor et al., 2008) or for vehicle models (Barros et al., 2017, 2016; Brackston et al., 2016; Littlewood and Passmore, 2012; Schmidt et al., 2015; Wassen et al., 2010).

The aerodynamic drag of bluff bodies can be reduced by avoiding velocity fluctuations. With two-dimensional flow around bluff bodies, the fluctuations in the wake in the form of the Kármán vortex street are strong sources of drag. For example, Pastoor et al. (2008) and Bock et al. (2016) showed that by influencing these coherent structures, the fluctuations can be attenuated and the base pressure can be increased by up to 40 %. The aerodynamic drag of bluff bodies mainly consists of the pressure difference between the front and the base (thus, pressure drag). Hence, the increase in the base pressure for bluff bodies can be transferred to the pressure drag to a similar extent. In many flows around bluff bodies, a significant portion of the aerodynamic drag is also to be expected in the velocity fluctuations. This physical connection can be shown in a variety of independent investigations. Balachandar et al. (1997) represented this through the integral influence variables on the base pressure starting from the separation streamline. In the work of Bock (2019), this was described by the effect of a potential vortex in the wall-bound vortex transport equation. In particular, vortex shedding is to be regarded as a fluctuating movement. Pastoor et al. (2008) showed this by using the example of the effect of vortex as a pressure sink on a wall in the form of the antisymmetric tensor in the Poisson equation of pressure.

To estimate the possibilities of aerodynamic drag reduction by influencing velocity fluctuations, prediction methods are required. In many cases, it can be advantageous to suppress the fluctuations in the wake, increase the base pressure and thus reduce the aerodynamic drag of the bluff bodies. However, the achievable base pressure is not known a priori. Only such a value would make it possible to classify the results of a flow control approach or assess its profitability. The necessity of determining the maximum possible base pressure was already addressed by Choi et al. (2008) in a review on the flow control on bluff bodies. Until now, there has been no work that makes direct reference to this question for an estimation of the minimum possible aerodynamic drag. However, various methods have been documented that use the separation velocity to calculate the base pressure. Documented methods for estimating and determining the base pressure on bluff bodies provide helpful approaches. Through this, a method could be developed to calculate a base pressure that can be expected for drag reduction approaches that avoid velocity fluctuations.

The base pressure of bluff bodies can be increased by measures of flow control, in either passive or active manner. Active flow control is very promising because of the advantages compared to passive measures. One example of such an advantage is the possible much smaller dimensions. However, active flow control needs a power input. To be able to estimate the allowable power consumption of an active flow control implementation it would be beneficial to know the possible power savings, e.g. drag reduction or base pressure increase. This paper suggests and investigates an approach to estimate the base pressure that would occur, when velocity fluctuations in the wake would be eliminated for example by active flow control. To compare the suggested estimation with actual achieved base pressures a flow around a bluff body without and with active or passive flow control is considered. The passive measure is represented by a splitter plate and the active measure by a Synthetic Jet actuation. The velocity fluctuations in the wake and therefore the low base pressure in this natural flow configuration is dominated by the Kármán vortex street. The behavior of this phenomena depends on the shape of the boundary layer at the separation edge. In addition, the suggested estimation for the possible base pressure increase is therefore considered for 3 different, artificially thickened boundary layers.

The Bernoulli equation provides a starting point for a base pressure estimation through the velocity. According to the Bernoulli equation, local pressure depends on local velocity. This applies to all attached flow areas. Thus, the pressure based on the local velocity along the boundary streamline is present at the limit of the dead water as a boundary condition. Hence, in a rough way the base pressure in the dead water can be determined via the velocity at the point of flow detachment.

The application of the velocity at the separation edge in an empirical base pressure representation yields good results. Roshko (1954) proposed determining the base pressure $c_{pb}$ via $c_{pb} = 1 - k$. The computation of k is based on the velocity of free streamlines at the point of detachment $u_s$ (with $k = (u_s / u_\infty)^2$ and $u_\infty$ the free stream velocity). Roshko (1954) used a theoretically determined separation velocity $u_s$ in his investigations. The considerations related to k show that this parameter describes the bluffness of a body. Bluffness is the ratio of the thickest diameter of dead water d' to the diameter d of the body. The determination of the separation velocity proposed by Roshko (1954) thus has a degree of freedom for the empirically determined influence of diffusion.

Compared to the measurement data, it can be shown that reasonable values of the base pressure can be determined for different bluff bodies if realistic values are chosen for k.

An example that directly employs the separation velocity in a calculation for the base pressure with the panel method does not show good agreement with the measured values. In a numerical simulation based on the methods of potential theory, Grün (1996) used the separation velocity to determine the total pressure loss ($c_{p,t} = 1 - (u_s / u_\infty)^2$) at the edge of the dead water area. However, the calculated base pressures are higher than the measured ones.

Nevertheless, a good correlation of pressure in the dead water and the velocity along the boundary streamline justifies the relationship between the separation velocity and base pressure. Maull & Young (1973) showed that Bernoulli's equation applied to the velocities along the boundary streamline in the shear layer can well reproduce the pressure in the same location. They also compared several experimental works of a flow over a backward facing step. These observations also show that, especially in the backward facing step flow, the pressure downstream of the base is relatively constant for more than half the dead water length.

The use of the separation velocity combined with additional variables in a semi-empirical model for aerodynamic drag correlates well with the measured values. In a study by Craze (1977), the aerodynamic drag is determined by integrating the diffusion term of the Navier-Stokes equation in the dead water volume and by estimating the characteristic quantities contained therein. This leads to a formulation that determines the aerodynamic drag and consequently the base pressure from the separation velocity. In this model for a disc in crossflow, however, additional terms are used for the influence of the empirically determined diffusion influences, such as the bending and the angle of the shear layer. Compared to some experimental data, the model provides good agreement for the drag coefficients.

Tanner (1973) proposed a more comprehensive model that is based on similar principles for calculating the base pressure of wedge flows with a splitter plate. Tanner's (1973) model is based on the following boundary conditions:

- The velocity along the separation streamline remains constant from the point of detachment to the downstream position of the near wake vortex center. Thus, it makes use of the abovementioned insight of Maull & Young (1973).
- The model requires highly resolved velocity profiles (one velocity profile in the center of the vortex and one in the wake behind the dead water) as a basis for determining the separation streamline.

The results from Tanner (1973) show that there is a strong influence of the separation velocity $u_s$, the velocity profiles (i.e., diffusion and turbulence), and the geometry (flow angle at separation) on the base pressure. A good estimation of the base pressure for different Mach numbers in the range of $0.1 < Ma < 0.8$ and wedge angles (flow angle at separation) from 0° to 90° is obtained. A problem for the generalization of this approach and a transfer to other flows is that realistic velocity profiles are required at the mentioned points. Since the reported observations of Tanner (1973) took place in a two-dimensional flow, the influence of three-dimensional effects could also be added.

The models presented for calculating the base pressure from the velocity at the separating edge motivate the use of this relationship. These models provide good results, especially for two-dimensional flow around a bluff body with a splitter plate. Based on these findings, the following section will develop a methodology that allows for the influence of the velocity fluctuations on the base pressure to be estimated. In section 3, experiments for the application of the proposed method are presented. The application of the method is examined in section 4. Finally, section 5 describes the results from the perspective of the applications of the proposed method.

## 2. ESTIMATION OF BASE PRESSURE

This work follows a facile approach, which determines the pressure level in dead water via the Bernoulli equation at the free streamline. The determination of fluctuation energy, i.e., the energy of the present turbulence, in addition to the involved coherent structures themselves, is a difficult undertaking, for which no approaches have yet been documented. For this reason, a different approach is taken here.

### 2.1 Assumptions for the estimation of the potential base pressure increase

It is assumed that a base pressure can be determined that would be approached if all velocity fluctuations were eliminated. This would correspond to a hypothetical case of a laminar bluff body flow without the occurrence of coherent structures. In section 1, previous works were presented that determined the base pressure from the influence of the velocity at the separation edge. All approaches of these methods use further terms as influencing variables to determine the base pressure. Based on the mentioned findings in section 1, the following assumptions for modeling are made.

The dead water area is considered under the following conditions:
1. The Bernoulli equation (e.g., $c_p = 1 - u^2 / u_\infty^2$) applies to the relation between the velocity u and the pressure coefficient $c_p$ outside the dead water area.
2. The curve of the separation streamlines in the wake determines the velocity at the dead water limit (cf. Maull & Young (1973)).
3. The separation streamline is located in a shear layer. Therefore, the pressure perpendicular to this line is considered to be constant.
4. The pressure level inside the dead water is considered to be predetermined at the separation streamline. For some flows (cf. backward facing step Tanner (1973), as well as Maull & Young (1973)), the velocity along the separation streamline (up to the downstream position of the vortex center) in the vicinity of the base is constant. This assumption is justified in view of the typical streamlines behind the separation edge, which initially undergo hardly any redirection. The velocity along the separation streamline is therefore the decisive influence on the pressure level in the dead water area.
5. The pressures inside the dead water and especially towards the base are then determined by internal (in the dead water) gradients (vortex sinks).

As described in section 1, the base pressure depends on the velocity at the separation edge. The velocities at characteristic boundary points between the separation edge and the free stagnation point could be used to determine the pressure boundary conditions, which could then be used to determine the initial pressure level. The works that are summarized by Tanner (1973) have done just that. For a wedge with a splitter plate this led to satisfactory results. However, there are still deviations. From points 1-5 above, it can be concluded that the base pressure is strongly influenced by the velocity at the separation edge or the total pressure loss at the separation edge. In practice, this is proven by the drag-reducing effect of boat-tailing and diffusers for passenger cars. These methods reduce the velocity at the separation edge and lead to an elevated base pressure (cf. Hucho (2013)).

Further influences on the base pressure are mainly due to velocity fluctuations in the dead water. Diffusion processes occur in the dead water, and the influence of these processes depends on velocity fluctuations, such as turbulence and coherent structures.

In Roshko's (1954) approach, implicit diffusion covers velocity fluctuations, vortex structures and turbulence. However, Roshko's (1954) model only applies a theoretical separation velocity for the base pressure computation. These theoretically determined separation velocities were not compared to the actual velocity at the separation edge within the scope of these investigations. Thus, the theoretical values may be much higher than the actual values. This means that all the effects of diffusion may be reflected in the separation velocity in this model.

**2.2 Connection between the base pressure and velocity fluctuations in the wake**

The consideration of diffusion as an influencing factor for the base pressure essentially leads to velocity fluctuations. In connection with the base pressure estimation and the influencing variables, the considerations of Balachandar et al. (1997) on the velocity fluctuations in the dead water area of a bluff body lead to helpful insights. If the flow around a bluff body is averaged over time, the dead water forms a closed domain with the base, and no streamline of the outer flow penetrates this domain. This domain is also limited by the separation streamline, which is located in the shear layer. The momentum balance of this boundary with the base surface represents the transfer of forces between the flow and the bluff body, which constitutes the proportion of the pressure drag. This momentum balance is shown in equation (1). The balance results from the stationary consideration along the boundary of the separation streamline $\Omega$ and the base surface. Since no streamline penetrates the separating streamline, the balance is without convection terms from the momentum balance (Navier-Stokes equation). Viscous friction is neglected due to the significantly higher turbulent viscosity. The sum of all base pressure forces is thus formed from two integrals along the separation streamline: the pressure coefficient $c_p$ and the time averaged velocity fluctuations $\langle u'_x u'_i \rangle$. The first integral is formed with the perpendicular direction to the flow $n_x$ and the second with each perpendicular direction $n_i$ for each velocity fluctuation component $u'_i$ in i (x, y, z) direction. The influence of the pressure (and thus the velocity in the potential flow or the region, which is free of rotation) from the separating edge along the separation streamline has already been discussed in subsection 2.1. The consideration from equation (1) shows a direct correlation of the velocity fluctuations to the base pressure. It also shows the weighting of velocity fluctuations as a sink of the base pressure. Furthermore, this analysis shows that the velocity fluctuations in the shear layer are included in this balance. At this location, they often represent the highest values in the flow field. This motivates the consideration of the velocity fluctuations in the wake and in particular structures of velocity fluctuations on the basis of energy-bearing vortex structures, in order to better understand the variables that influence aerodynamic drag.

$$\int_{\text{base surface}} c_p \, ds = \int_{\Omega} c_p \cdot n_x \, ds + \int_{\Omega} \frac{2}{u_\infty^2} \cdot \langle u'_x u'_i \rangle \cdot n_i \, ds \qquad (1)$$

The velocity gradients caused by the diffusion processes may induce an upstream reaction to the velocity at the separation edge. From the momentum balance, the terms for acceleration and curvature (e.g., the radial pressure equation) must be balanced locally with the pressure gradient and the diffusion terms. Therefore, feedback of the diffusion processes in the wake to the upstream velocity at the separation edge or its pressure level is also possible. However, in many cases the pressure determined by the specific separation velocity is higher than the actual base pressure (cf.

Grün (1996)). Motivated by this understanding and that of the works of Roshko (1954), Maull & Young (1973) and Grün (1996), the feedback of diffusion processes is neglected in this work.

### 2.3 Estimation of the potential base pressure increase

The base pressure essentially depends on the velocity at the separation edge. However, this effect on the base pressure is further corrupted by velocity fluctuations in the dead water. Together with the conditions described here, the following hypothesis is established, tested and applied in the context of this work. A base pressure coefficient or a total pressure loss portion can be determined from the velocity at the separation edge $c_{pb,s}$. This is defined as:

$$c_{pb,s} = 1 - \left(\frac{u_s}{u_\infty}\right)^2 \tag{2}$$

This base pressure coefficient represents a total pressure drop that is lower than the actual total pressure drop. In other words, the base pressure coefficient from the velocity at the separation edge $c_{pb,s}$ is higher than the actual pressure coefficient: $c_{pb}$. The difference is: $\Delta c_{pb,s} = c_{pb,s} - c_{pb}$.

In this sense, the base pressure, which is calculated from the velocity at the separation edge, is regarded as a limit value that is reached if all the velocity fluctuations are suppressed. This can be argued as follows. A high energy content of the velocity fluctuations in the dead water results in a large pressure differential $\Delta c_{bb,s}$. The velocity fluctuations therefore cause a base pressure that is lower than $c_{pb,s}$. The pressure difference $\Delta c_{bb,s}$ mainly represents the influence of the energy of the velocity fluctuations due to turbulence and coherent structures on the base pressure. It is implicitly assumed that the diffusion processes caused by turbulence and coherent structures strongly outweigh all others (pure molecular friction). Furthermore, the feedback of the diffusion processes (or velocity fluctuations) to the velocity at the separation edge is neglected. If the amount of specific coherent structures in the velocity fluctuations is dominant, avoiding these coherent structures can increase the base pressure by a large component of the pressure difference $\Delta c_{pb,s}$. In this sense, the pressure difference $\Delta c_{bb,s}$ can be regarded as the potential in flow control applications.

For verification, velocity profiles along the free streamline and the development of the shear layer thickness along the dead water are considered to be indicators for upstream feedback via diffusion. The free streamline is used instead of the separation streamline here, and it represents the streamline beginning with a wall distance equal to the momentum thickness $\delta_1$ from the separation edge. The relationship between the base pressure and velocity at the separation edge works well in the two-dimensional flow around a bluff body with a splitter plate. At other flows, further influencing factors could gain importance. These influencing factors should have an effect on the diffusion in the shear layer, and thus on the development of the velocity along the free streamline and the shear layer thickness. A comparison of these variables between the two-dimensional flow around the bluff body with a splitter plate and other flows should therefore be used here as an evaluation criterion. This criterion accounts for the significance of the relationship between the base pressure and the separation velocity.

The shear layer thickness $\delta_S$ is described by the maximum, minimum velocity $u_{max}$, $u_{min}$ along the profile perpendicular to flow direction and by the velocity gradient $\partial \bar{u}/\partial y$ at the location of the free streamline $y_s$ for each point in flow direction as:

$$\delta_S(x) = \frac{u_{max} - u_{min}}{\left|\frac{\partial \bar{u}}{\partial y}\right|_{y_s}} \quad (3)$$

## 3. EXPERIMENTAL SETUP OF A BLUFF BODY IN TWO-DIMENSIONAL FLOW

To examine the proposed approach, the base pressures and velocities at the separation edges were measured for two configurations of bluff body flows. The experiments were conducted on a bluff body in two-dimensional flow with and without a splitter plate. All experiments were conducted in the Göttingen-type wind tunnel of the Institute for Internal Combustion Engines and Automotive Engineering of the University of Stuttgart. The wind tunnel is equipped for measuring vehicle models on a scale of 1:4 and 1:5. The nozzle outlet has a height of $h_n = 1.055$ m and a width of $b_n = 1.575$ m, and the test section length is $l_{ts} = 2.585$ m. The main characteristics of the wind tunnel are described by Wiedemann & Potthoff (2003).

### 3.1 Wind tunnel setup

To prepare the test section for measuring a two-dimensional flow, the symmetry boundary conditions of the wind tunnel were adapted by using a ceiling plate in the upper shear layer between the nozzle and collector. Figure 1 (top) schematically shows this configuration with the nozzle outlet on the left, the collector on the right and the model of the bluff body in the middle. Like the floor, the ceiling panel extends from the edge of the nozzle to the collector of the wind tunnel. The ceiling plate and floor thus represent the end plates of the configuration. The bluff body model has a length of $l = 1.077$ m and a height of $h = 0.1$ m. Like the nozzle height, the width of the model $w = 1.055$ m is between the end plates. The leading edge of the model is located at a distance of 0.321 m from the nozzle exit plane. The nose of the model is designed semi-elliptically with a half axes ratio of 4. The x and y-axes of the configuration were aligned with the length and height directions, respectively. The origin of this coordinate system is defined as the center of the base, as marked with a ⊕ in Figure 1. The z-axis therefore runs along the symmetry axis of the model. The green areas in Figure 1 illustrate the light sheet that is spanned for PIV measurements (see section 3.3). To change the thickness of the boundary layer at the separation edge, a film coated with a high roughness (sandpaper) was successively applied by gluing on the surface behind the elliptical nose. The boundary layer thickness was thus varied in three different steps. The Reynolds number varied between $Re_h = \{45000, 65000, 90000\}$ by the velocities $u_\infty = \{7, 10, 14\}$ m / s, respectively. The degree of turbulence at the nozzle outlet at these velocities is less than $Tu < 1.3$ %. The boundary layer thickness was varied through the sandpaper at Reynolds number $Re_h = 45000$, where three boundary layer thicknesses $\delta_1 / h = \{1.6, 2.7, 3.5\} \cdot 10^{-2}$ (without a splitter plate) and $\delta_1 / h = \{2.7, 2.5, 2.4\} \cdot 10^{-2}$ (with a splitter plate) were investigated. To investigate the case of two-dimensional flow without a Kármán vortex street by means of passive flow control, a $l_{tp} = 4.5 \cdot h$ long and $h_{tp} = 0.02 \cdot h$ thick splitter plate aligned in the direction of flow was installed in the middle of the base height. This configuration is depicted in Figure 1 (bottom). Similarly, Synthetic Jets were applied as active flow control alternative to attenuate the Kármán vortex street. Splitter plate and Synthetic Jets were applied to investigate the effect on base pressure increase in comparison with the proposed method for base pressure estimation for suppressed wake velocity fluctuations. The roughness elements were considered to study the effect of different boundary layers on the estimation of the proposed method.

The reference velocity $u_\infty$ is the calibrated wind tunnel velocity without the model (but with end plates). Here, the interference effects were taken into account in accordance with the methods proposed by Mercker et al.(1997) and Mercker & Wiedemann (1996). These formulations take the interference effects of nozzle blocking and deflection, collector blocking and pressure gradients into account via models of potential theory. These methods were transferred to the existing two-dimensional flow. However, the procedure follows the same approach. The correction factors for $c_q = 1.0547$ for the dynamic pressure and $c_u = 1.027$ for the velocity were determined from this method. The correction factors are then used in the sense of global values for the reference velocity $u_\infty$ and for quantitative comparisons of the base pressure. A local correction would be necessary for field data. However, this is only necessary for a quantitative comparison of the absolute values and is therefore not used here.

A Synthetic Jet was generated at the trailing edge of the model by a net-zero mass flow actuator. The Synthetic Jet is generated via an orifice at the separation edge. The orifice is oriented at an angle of 45° to the free flow, resulting in the outflow direction shown in Figure 1. Different velocity amplitudes $u_a / u_\infty$ and dimensionless frequencies $Sr_a = f_a h / u_\infty$ of the Synthetic Jet can be realized as described in Bock et al. (2016) and Bock (2019). Some particular cases of these works are considered here regarding the resulting base pressure.

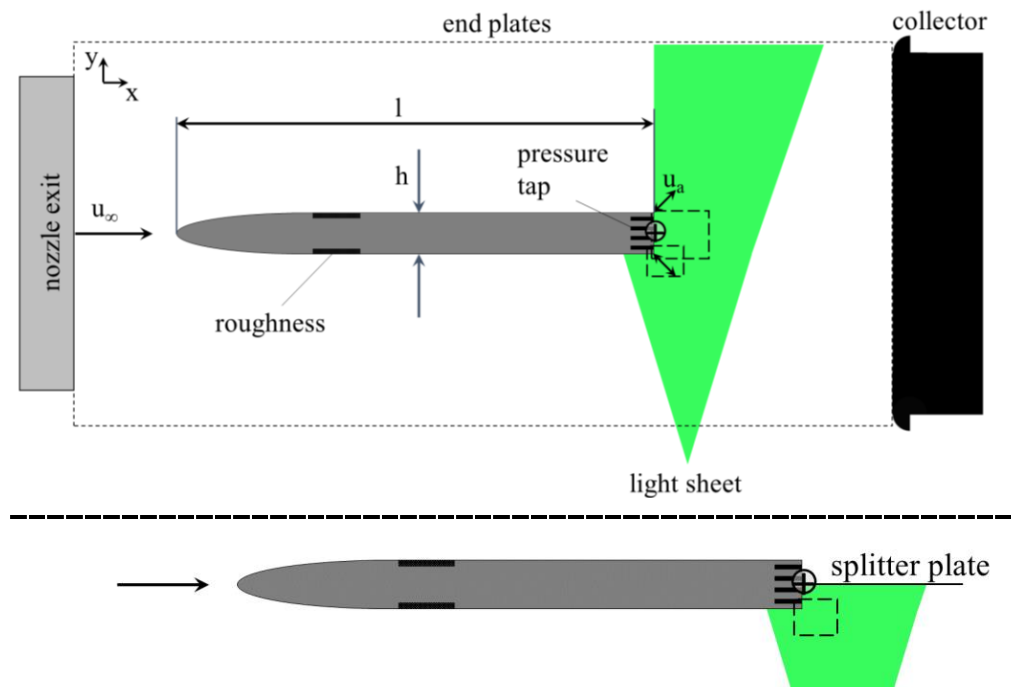

**Figure 1**: Experimental setup for two-dimensional bluff body flow. Top: Two-dimensional setup in a wind tunnel. Bottom: Two-dimensional setup with a splitter plate.

### 3.2 Base pressure measurements

The base pressures were measured and time-resolved at 84 measuring points using the pressure taps marked in Figure 1. The position of the pressure taps were in vertical direction and in perpendicular direction to the view in Figure 1, respectively:

y / h = {-0.43, -0.35, -0.27, -0.21, -0.15, -0.09, -0.03, 0.03, 0.09, 0.15, 0.21, 0.27, 0.35, 0.43}
z / h = {-3.75, -2.25, -0.075, 0.075, 2.25, 3.75}.

Each pressure tap has an internal diameter of $d_a = 0.7$ mm and is connected to an Esterline ESP-64HD pressure transducer by a 0.5 m long hose with an internal diameter of 1.4 mm. The pressure transducers are calibrated and have temperature compensation. The operating range of the pressure transducer is up to 6.9 kPa with an accuracy of 0.05 %, and the sampling rate is 250 Hz. To take the dynamic effects of the system's hose and the measuring volume of the pressure transducer into account, a dynamic correction of the measured pressure values according to Bergh & Tijdemann (1965) was applied. Each pressure measurement took place over a total duration of 65.5 seconds. This corresponds to approximately $2^{14} = 16384$ measured values or time instants per pressure tap. The measured pressure values are represented as dimensionless values $c_p = \Delta p / q_\infty$ by the pressure difference $\Delta p = p_S - p_\infty$ to the plenum pressure of the wind tunnel $p_\infty$ in relation to the dynamic pressure $q_\infty = 0.5 \, \rho_\infty \, u_\infty^2$. The measured taps in z direction were spatially averaged at the same vertical position (y = constant).

### 3.3 Measurements of the flow field

Particle Image Velocimetry (PIV) was used to observe the flow around the separation edge and the wake. For this purpose, an Nd:YAG laser with a fixed frequency; thus, a sampling rate of 10.5 Hz, a wavelength of 532 nm, and a maximum energy of 850 mJ per pulse was used to span a light sheet in the z-plane around and behind the trailing edge of the model. The thickness of the light sheet was approximately 2 mm. The incoming flow was mixed with droplets of Di-Ethyl-Hexyl-Sebacat (DEHS) with a diameter of approximately 1 µm. Using a two imager sCMOS with a resolution of 2560 x 2160 pixels, the images were recorded in a mono-PIV configuration. This means that two different fields of view are available. A camera recorded the area around the separation edge with a 200 mm lens, resulting in a field of view of 1.03 x 1.04 h. The second camera was directed at the wake with a 135 mm lens and had a field of view of 1.47 x 1.49 h. The time delay of the double images was chosen to obtain a maximum pixel offset of 6. For the temporal averaging, 700 snapshots were recorded. The velocity vectors were calculated with an interrogation window of a 32 x 32 pixel with a 75 % overlap. In the small and large field of view the resolution is 0.0032 h and 0.0046 h, respectively. No smoothing or filter was used in the evaluation.

### 3.4 Measurement of the velocity profiles at the separation edge

The measurements in the small field of view included the velocities around the separation edge (2 mm upstream of the edge) with a resolution of 0.0032 h (0.32 mm). Due to the optical accessibility for the measurement, only the profile at the lower edge was used. Symmetrical conditions between the upper and lower edge can be assumed. From this field, the boundary layer profile as shown in Figure 2 with the wall distance on the ordinate and the velocity on the abscissa can be extracted. In the picture, the individual measuring points are marked with a +. The velocity profile shows that the outflow velocity is greater than $u_\infty$ due to local blocking effects.

The dimensionless curves only differ between cases with and without a splitter plate. However, in the case without a splitter plate, the downstream effects of the gradients in the wake region on the separation edge are so strong that an overshoot occurs in the profile, with the maximum velocity at the position y ≈ $\delta_{99}$. In all other cases, the boundary layer thickness $\delta_{99}$ was defined by the first digit coming from the wall (y = 0), where the velocity is u > 99 % of the maximum velocity of the profile. The boundary layer parameters of displacement thickness $\delta_1$ and momentum loss thickness $\delta_2$ were determined by integration in the range $0 \leq y \leq \delta_{99}$, assuming piecewise linear sections between the points shown in Figure 2. The velocity at the separation edge for the estimation of the proposed potential for a base pressure increase was defined as the maximum velocity in the profile, which is in proximity to the position of $\delta_{99}$.

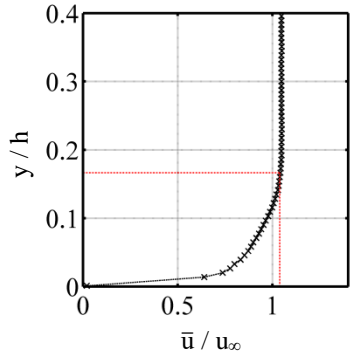

**Figure 2:** Boundary layer at the separation edge at Reynolds $Re_h$=45000 with a splitter plate.

## 4. RESULTS
In this section, the proposed method for a possible base pressure increase by avoiding velocity fluctuations is applied. Different cases of bluff bodies in two-dimensional flow are examined, and the quality of the prediction is assessed.

### 4.1 Flow field
The flow fields in the wake of the two-dimensional flow around a bluff body are shown in Figure 3 and Figure 4. Figure 3 shows the wake of the flow without a splitter plate and Figure 4 shows a section of the wake with a splitter plate. In these two representations, the rear section of the bluff body is depicted as a gray area. These figures show the streamlines of the wake and the coloring of the velocity fluctuations u'. The free streamlines are plotted as thick lines, and they are defined at a starting point with distance of $\delta_1$ to the separation edge of the bluff body. The focus of the stationary vortices are marked with a black +. In the two-dimensional flow without a splitter plate in Figure 3, the strongest fluctuations arise downstream of the focuses of the vortices around the saddle point (at x / h ≈ 0.8). The depicted section of the two-dimensional flow with a splitter plate in Figure 4 is smaller than the section for the flow without a splitter plate. Nevertheless, comparing the streamlines, especially the free streamline in the range 0 < x / h < 0.4, indicates a faster convergence to the y = 0 axis or the saddle point for the flow without a splitter plate. Thus, streamlines have stronger curvature without a splitter plate. This means there are stronger gradients in the wake. From the comparisons of these flows it can also be observed that the maximum velocity fluctuations in the comparable range (0 > x / h > 0.4) are higher for the flow without a splitter plate (u' / $u_\infty$ ≈ 0.3) than with a splitter plate (u' / $u_\infty$ ≈ 0.2). This correlates with the faster convergence of streamlines from the two shear layers without a splitter plate.

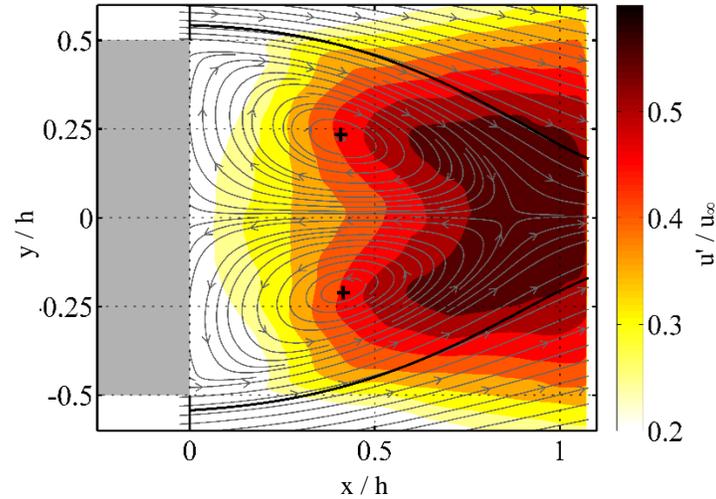

**Figure 3:** Flow field in the wake of the bluff body in two-dimensional flow at $Re_h = 45000$. The coloring shows the magnitude of the velocity fluctuations.

To estimate the possible base pressure increase, the velocity profile at the separation edge was determined. The local boundary layer profiles are used to estimate a minimum base pressure without the influence of velocity fluctuations (via the hypothesis presented in section 2) from the velocity at the separation edge. For this purpose, the boundary layer profile was extracted from the flow field at the separating edge, as described in 3.1.

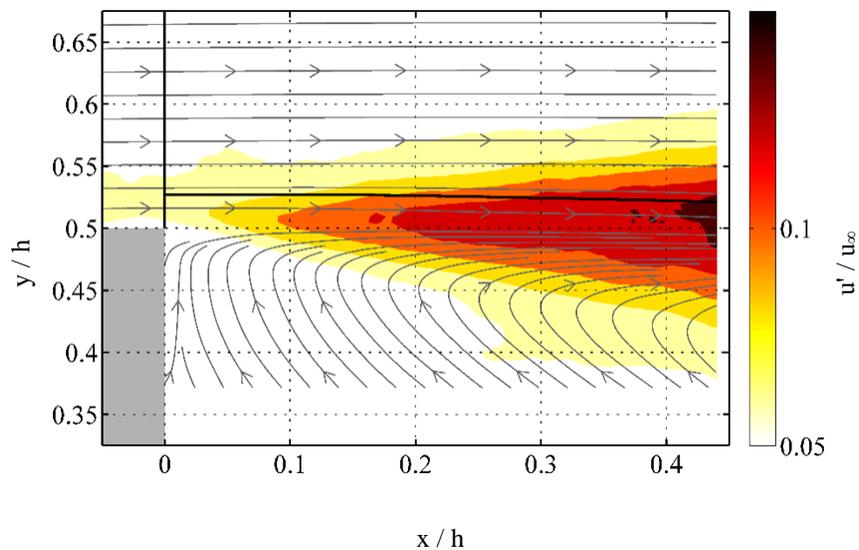

**Figure 4:** Section of the flow field in the wake of the bluff body in two-dimensional flow with a splitter plate at $Re_h = 45000$. The coloring shows the magnitude of the velocity fluctuations.

### 4.2 Velocity profiles at the separation edge

Without a splitter plate, the velocity profiles are under a stronger influence of the downstream gradients. Figure 5 shows the boundary layer profiles of some cases. The image shows the dimensionless velocity $\bar{u} / u_\infty$ distribution on the ordinate versus the wall distance on the abscissa. One boundary layer profile with a splitter plate and three without splitter plates with different boundary layer thicknesses ($\delta_1 / h$) (all with Reynolds number $Re_h = 45000$) are shown. It should be noted that the profiles for the other measured Reynolds numbers ($Re_h = 70000 / 90000$) do not differ (comparing the integral values in Table 1). However, as expected the boundary layer profiles at different boundary layer thicknesses show clear differences. In comparison to the case with the splitter plate, it is noticeable that there is an overshoot without a splitter plate. Above the overshoot, these profiles begin to approach the profile with the splitter plate. This shows that at this point the velocities differ due to local gradients. The boundary layer with a splitter plate resembles a flat plate boundary layer without pressure gradients. Thus, it can be assumed that the boundary layer in the configuration without a splitter plate is more influenced by downstream gradients than by other factors.

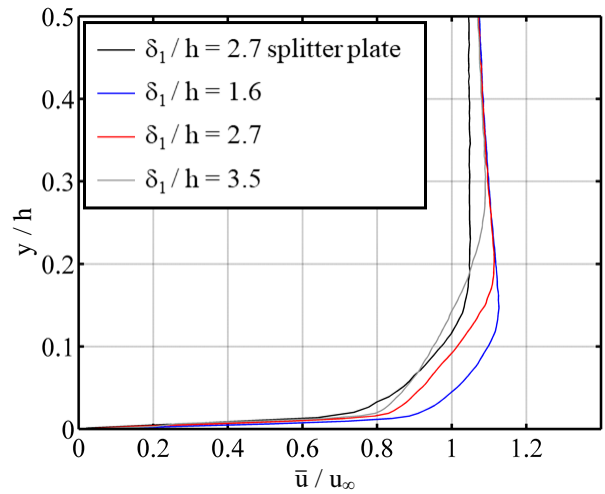

**Figure 5:** Boundary layer profiles at the separation edge for different boundary layer thicknesses at Reynolds number $Re_h = 45000$.

### 4.3 Base pressure estimation

The integral boundary layer thicknesses differ when using the splitter plate. The velocity profiles were used to determine the velocities at the separation edges $u_s$ and the boundary layer properties. The boundary layer properties are the displacement thickness $\delta_1$, the shape factor $H_{12}$ between displacement thickness and momentum thickness and the boundary layer thickness $\delta_{99}$. These values are listed together in Table 1 with the temporally measured and spatially averaged base pressures $c_{pb}$. First, the form factors of all cases are considered, suggesting that it is a fully turbulent boundary layer before detachment. The boundary layer thicknesses $\delta_{99}$ and in particular $\delta_1$ and $\delta_2$ hardly differ for different Reynolds numbers $Re_h$. A striking feature in the comparison of the boundary layer values results from the fact that the boundary layer thicknesses (especially $\delta_1$) with a splitter plate are higher than without a splitter plate. A slight difference can also be seen between the different boundary layer thicknesses $\delta_1$ in the case without a splitter plate. The comparison of these values is a quantitative characterization of the qualitative observations described in section 4.2 from Figure 5.

**Table 1:** Properties in the boundary layer and estimation of the achievable base pressure on a bluff body in two-dimensional flow. *[1] $Sr_a = 0.17$ & $u_a / u_\infty = 1.3$, *[2] $Sr_a = 0.24$ & $u_a / u_\infty = 1$, *[3] No determination of the flow field and boundary layer.

| $Re_h \cdot 10^4$ | $\delta_1 / h \cdot 10^{-2}$ | $H_{12}$ | $\delta_{99} / h \cdot 10^{-2}$ | $u_s / u_\infty$ | $c_{pb,s}$ | $c_{pb}$ | $(c_{pb} - c_{pb,s}) / c_{pb}$ | $c_{pb}'$ |
|---|---|---|---|---|---|---|---|---|
| 4.5 | 1.6 | 1.6 | 12.6 | 1.14 | -0.30 | -0.48 | 0.37 | 0.09 |
| 4.5 | 2.7 | 1.5 | 17.7 | 1.12 | -0.26 | -0.44 | 0.41 | 0.09 |
| 4.5 | 3.5 | 1.5 | 25.0 | 1.09 | -0.19 | -0.46 | 0.58 | 0.09 |
| 4.5*[1] | *[3] | *[3] | *[3] | *[3] | *[3] | -0.35 | - | 0.18 |
| 4.5*[2] | *[3] | *[3] | *[3] | *[3] | *[3] | -0.53 | - | 0.21 |
| 6.4 | 1.8 | 1.5 | 13.5 | 1.13 | -0.28 | -0.49 | 0.43 | 0.08 |
| 7.0 | 1.8 | 1.6 | 13.5 | 1.14 | -0.29 | -0.49 | 0.41 | 0.08 |
| with a splitter plate | | | | | | | | |
| 4.5 | 2.7 | 1.7 | 16.7 | 1.05 | -0.10 | -0.10 | - | 0.04 |
| 6.4 | 2.5 | 1.4 | 16 | 1.05 | -0.10 | -0.17 | 0.42 | 0.02 |
| 7.0 | 2.4 | 1.4 | 15.4 | 1.05 | -0.11 | -0.17 | 0.36 | 0.02 |

Consistent with the observations of the boundary layers, the maximum velocities for cases with and without splitter plates have the greatest difference. For different Reynolds numbers $Re_h$ the maximum velocities $u_s / u_\infty$ hardly differ in the boundary layer. These trends are also consistent with the measured base pressure coefficients. The changes in the measured base pressure coefficient $c_{pb}$ and $c_{pb,s}$ determined from the velocity at the separation edge for different Reynolds numbers $Re_h$ behave approximately the same.

In the cases considered (except with a splitter plate at $Re_h = 45000$) in Table 1, relative increases in the base pressure (($c_{pb}$ - $c_{pb,s}$) / $c_{pb}$) between 36 % and 58 % are possible according to the hypothesis established here based on the separation velocity $u_s$.

In fact, higher base pressure coefficients can be achieved by using the splitter plate than what is predicted with the hypothesis. This is indicated by a comparison of the base pressure coefficients determined from the velocity at the separation edge (without a splitter plate) against the measured values (with a splitter plate). For cases with a splitter plate, the measured base pressures with a splitter plate are always higher. This means that by reducing the velocity fluctuations in the wake through the splitter plate, the base pressure is always higher than the base pressure coefficient determined by the velocity at the separation edge. This leads to the conclusion that higher base pressure coefficients can actually be achieved than those predicted with the proposed method. With the actual results of the active flow control approaches, base pressure increases up to 28 % are achieved. However, the estimated values have not exactly been reached yet (compared to Pastoor et al. (2008), Bock et al. (2016)).

The increase in the base pressures relates to a reduction of the velocity fluctuations of energy bearing coherent structures. The fluctuations in the base pressure of all flows without a splitter plate are at the same level. With a splitter plate, the fluctuations of the base pressure are reduced. The fluctuations of the base pressure $c_{pb}'$ with a favorable active control do not reflect this at first glance. However, this requires a breakdown of the fluctuations and their sources. Bock (2019) showed that the fluctuations resulting from the large coherent structures in the flow field are reduced when excitation is favorable and vice versa when excitation is not favorable. This is consistent with the hypothesis of reducing the base pressure when reducing the velocity fluctuations of energy bearing coherent structures.

The thicker boundary layers with a splitter plate can be explained by the stronger gradients in the case without a splitter plate. These are reflected in the overshoot of the velocity profile (see

Figure 5), which only occurs in cases without a splitter plate. Due to the Kármán vortex street, which occurs without a splitter plate, higher velocity fluctuations are caused in the shear layer and in the wake. This causes a stronger mixing in the flow and provides the stronger gradients. These strong gradients are confirmed when comparing the averaged flows (cf. section 4.1).

According to the hypothesis of this work, a higher base pressure is made possible for the thicker boundary layers by reducing the velocity fluctuations than those for smaller boundary layer thicknesses. Larger deviations in the estimation of the possible base pressure increases occur for different boundary layer thicknesses $\delta_1 / h$. The measured base pressure differs for these cases. However, the difference in the base pressure and in the velocity fluctuations is not very high. The relationship between the boundary layer thickness and the maximum achievable base pressure represented by the hypothesis does not appear to be justified at this point. This discrepancy cannot be definitively clarified with the available measurement data. Nevertheless, in the studies by Bock et al. (2016) and Bock (2019), a higher base pressure increase with thicker boundary layers was achieved by measures of active flow control. Furthermore, the results of Bock (2019) indicated that the shares of fluctuations of the most energetic coherent structures also differed for these cases.

According to the hypothesis, an additional increase in the base pressure could be achieved by further reducing the velocity fluctuations. Table 1 shows that the cases of maximum base pressure (e.g., with a splitter plate) are still subject to fluctuations. Furthermore, the base pressure from the estimation is lower than the measured base pressure for two of the three cases. This indicates that a base pressure increase is still possible with a splitter plate. However, these differences in the base pressure and the magnitude of the fluctuations are only small compared to the flow without a splitter plate. This suggests that the achievable base pressure increases are also small. Examples of active flow influence of the coherent structures at backward facing steps of Dahan et al. (2012), Gong (2015), Chun & Sung (1996) correspond to cases with a splitter plate. These confirm the statement that fluctuations of the dominant coherent structures in this flow can actually be reduced (Chun and Sung, 1996; Dahan et al., 2012) and that the base pressure increases by up to 4-20 % (Dahan et al., 2012; Gong, 2015). However, this also shows that these changes are small compared to the case without a splitter plate.

**4.4 Free streamline velocity profiles**
The flow with a splitter plate is used as a reference for observing the constant velocity along the free streamline to the downstream position of the vortex center. To check the assumed boundary condition of the constant velocity along the free streamline, the velocities along the free streamline were extracted from the data of the wake. The starting point for the streamline was the position with an orthogonal distance to the flow (y) direction corresponding to the displacement thickness $\delta_1$ from the separation edge. The flow case with the splitter plate corresponds to the case for which Tanner's (1973) model, mentioned in section 2, has been developed. These cases should serve as a reference for the sufficiently fulfilled boundary condition of constant velocity along the free streamline.

The velocity curves along the free streamline show severe changes without a splitter plate and can only be regarded as constant up to half the distance to the downstream center of the vortex. Figure 6 compares the velocity magnitude along the free streamline for the two-dimensional flow with and without a splitter plate in the wake. The curves of the velocity magnitude over the distance x / h from the base downstream for different boundary layer thicknesses (without a splitter plate) and for one boundary layer thickness with a splitter plate are shown. The profiles of the free streamline for different Reynolds numbers $Re_h$ through different inflow velocities are very similar and are not shown here. The most pronounced change in the slope can be observed in the downstream extend of the near wake. The velocities change in the range under consideration ($0 < x/h < 0.45$) from $0.9 > u/u_\infty > 0.6$ without a splitter plate, and from $0.8 > u/u_\infty > 0.75$ with a splitter plate. Without a splitter plate, the observed area covers the path from the base to the vortex center at $x/h = 0.4$. With a splitter plate, the velocity changes less than without. However, for $0 < x/h < 0.15$ there is a range that is the closest to a constant behavior for all cases, except for $\delta_1/h = 1.6 \cdot 10^{-2}$, with a tolerance of $u/u_\infty < 0.02$. As the distance from the separation edge increases, the velocity decreases slightly in all cases. Without a splitter plate, the drop in velocity is more abrupt. The strongest changes between the starting point and the end of the profile can be examined for the lowest boundary layer thickness. Thus, the flow without a splitter plate shows larger deviations to the assumption of the constant velocity of Tanner (1973) than the flow with a splitter plate. In the case of the lowest boundary layer thickness ($\delta_1/h = 1.6 \cdot 10^{-2}$), the changes along the free streamline are the strongest.

The deviation from the constant velocity along the free streamline to the downstream position of the center of the vortex correlate with an underestimation of the actually achievable base pressure by avoiding velocity fluctuations. A constant velocity along the free streamline to the vortex center is an indicator of how good the estimation of the saving potential from the separation velocity is. If the velocity curve is not constant, then the influence of diffusion may be strong. In an unfavorable case, this influence can extend very far upstream, and it could potentially influence the measured separation velocity itself. In the case without a splitter plate, due to the nonconstant velocity up to the center of the vortex, stronger repercussions of the diffusion in the wake on the separation edge are to be expected. This compromises the significance and the accuracy of the estimation of the possible base pressure by avoiding velocity fluctuations.

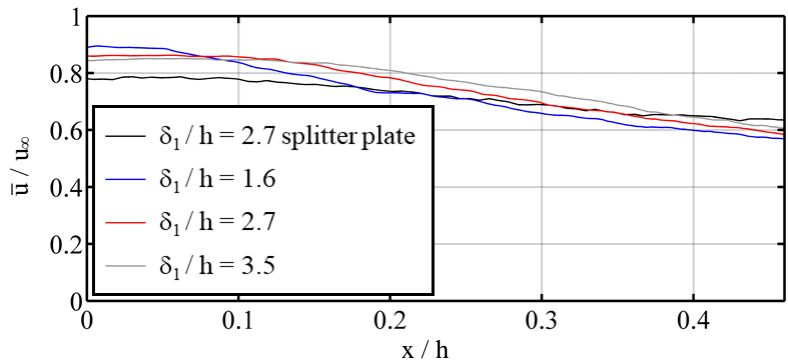

**Figure 6:** Velocity development along the free streamline in the wake of the bluff body in two-dimensional flow for different boundary layer thicknesses.

### 4.5 Development of shear layer thickness
The change in the thickness of the shear layer with the distance from the separation edge reflects the diffusive influences on the velocity curves along the free streamline, which reflects the estimation of the base pressure. Figure 7 shows the development of the shear layer thickness $\delta_S/h$ in the dead water. The shear layer thicknesses above the distance $x/h$ downstream from the base for different boundary layer thicknesses without a splitter plate and for one boundary layer thickness with a splitter plate are plotted. The initial values of the shear layer thicknesses approximately correspond to the boundary layer thickness (see Table 1). With a splitter plate, the shear layer grows with a constant gradient ($d\delta_S / dx \sim 0.2$). In the range $0 < x / h < 0.15$ the shear layer thickness changes only slightly for cases without a splitter plate or grows at most to the same extent as for the flow with a splitter plate. This is analogous to the velocity along the free streamline, which changes little in this range. At greater distances from the separation edge the shear layer grows nearly linear for the flow with splitter plate. For cases without splitter plate this growth is the more progressive the thicker the boundary layer.

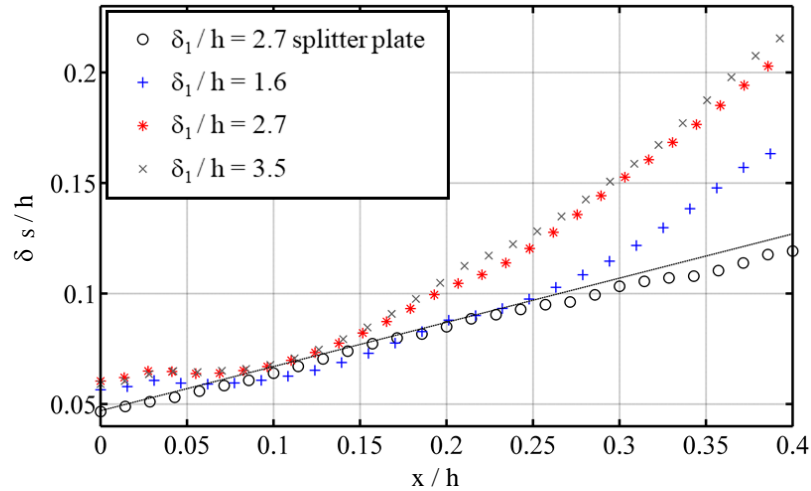

**Figure 7:** Development of the shear layer thickness in the wake of a two-dimensional bluff body ($d\delta_S / dx \sim 0.2$).

The development of the shear layer thickness does not explain all the differences in the base pressure estimation. However, the slope increases the least with the thinner boundary layer $\delta_1 / h = 1.6 \cdot 10^{-2}$. This case obviously has more diffusive losses compared to the base pressure than the others do. Hence, the consideration of shear layer thickness does not cover all influencing variables that are responsible for the difference in the prediction. In cases without a splitter plate, in the range $x / h < 0.05$ a further deviation from the linear increase in the shear layer thickness also occurs. This phenomenon could be an artifact of the response of the diffusion in the shear layer on the area around the separation edge.

The accuracy of the estimated potential for savings is reduced by the diffusion influences. The shear layer thickness with a splitter plate increases linearly with the distance from the separation edge. This corresponds to the typical development of a free shear layer and justifies the assumption of a constant velocity along this region in the free streamline for this flow.

Without a splitter plate, the shear layer thicknesses near the separation edge also initially increase almost linearly, or they do not change very strongly, which is in line with the velocity of the free streamline. However, with an increasing distance the shear layer thicknesses increase more and thus show a stronger influence of diffusion than with a splitter plate. This increase in the shear layer thicknesses is related to the intensity of the velocity fluctuations in the shear layer. Without a splitter plate, a Kármán vortex street (see Bock (2019)) occurs, which is characterized by the strong interaction of shear layers and the strong coherent structures. The stronger the potential by avoiding velocity fluctuations, the less accurate the prediction becomes.

# 5. CONCLUSION

To estimate the base pressure on bluff bodies, which results from the avoidance of diffusive losses in the form of fluctuating movements, the application of a formulation on the velocity at the separation edge was proposed. The calculation is based on the assumed constant velocity curve along the separation streamline. Along this curve, a constant velocity between the separation edge and the downstream position of the wake vortex in the main flow direction is assumed, as in the two-dimensional flow around a bluff body with a splitter plate. The proposed estimation and compliance with the boundary condition was conducted for a bluff body in two-dimensional flow with and without a splitter plate.

The quality of the estimation is different for different flows. The base pressure in the two-dimensional flow without a splitter plate is lower than it is in other observed flows. The velocity fluctuations and the shear layer thickness standing for diffusive influences increase much faster downstream of the separation. These effects lead to stronger deviations in the constant velocity along the free streamline. Hence, the assumptions made for the estimation of the base pressure by avoiding velocity fluctuations are not as well maintained as in the flow without a splitter plate. The determined estimates of the base pressure with avoidance of velocity fluctuations without a splitter plate give lower values than those actually achievable with a splitter plate. Hence, this prediction underestimates the achievable values and is therefore on the conservative side. The extent of underestimation depends on the effect of the diffuse losses. With an example case of the active flow control, the estimated base pressures can almost be reached even without eliminating all velocity fluctuations in the wake.

The savings potential must be considered in individual cases, as general statements are not possible on the basis of these results. The values of the base pressure determined via the separation velocity and the difference to the measured value considered here are consistent with the bluff body in two-dimensional flow with a splitter plate. This difference is always estimated as too small in two-dimensional flow, i.e., between with and without a splitter plate. This can be attributed to the repercussion of diffusion in the shear layer on the upstream separation edge. In other words, these repercussions reflect the nonlinear behavior of the flow, which in this case is due to the reaction of the change in the wake on the boundary layer and thus the velocity at the separation edge. However, underestimating the possible increase in the base pressure by avoiding velocity fluctuations in the two-dimensional flow cannot necessarily be generalized.

The proposed approach represents an up-to-date undocumented possibility for estimating the possible base pressure increase via control approaches, which can be applied to a multitude of flows around the bluff bodies. Thus, the approach is a helpful tool that can be used to estimate the possible savings due to active flow control of coherent structures. It should be noted that velocity fluctuations consist of different coherent structures and a remainder of velocity fluctuations, which instead represent disorderly fluctuations. Thus, a measure of the flow control aimed at certain coherent structures can never eliminate all velocity fluctuations. In this respect, the estimation of the possible base pressure by avoiding velocity fluctuations is to be regarded as a limit value. The accessibility of the required measured values (in the velocity field of the wake) for corresponding investigations is improving, especially with current optical measuring methods such as PIV and tomographic PIV, which are currently under development. A theoretical or empirical approach to consider the influence of diffusion is a promising improvement on the quantitative estimation.

**ACKNOWLEDGEMENTS**
The author acknowledges the support of the Institute for Internal Combustion Engines and Automotive Engineering

University Stuttgart where the experiments were performed. Furthermore, the support of the Friedrich-und-Elisabeth-BOYSEN-Stiftung is thankfully acknowledged for funding under grant BOY11-No.77. The author also warmly thanks Nils Widdecke, Timo Kuthada, Christoph Schönleber, Alexander Hennig, Max Tanneberger and Daniel Stoll for their fruitful discussions.**REFERENCES**